# ESTIMATING INDOOR CROWD DENSITY AND MOVEMENT BEHAVIOR USING WIFI SENSING

A PREPRINT

Syed Salman Alam
Department of Research
Elm Company
Riyadh, Saudi Arabia

Muhammad Al-Qurishi
Department of Research
Elm Company
Riyadh, Saudi Arabia

Riad Souissi
Department of Research
Elm Company
Riyadh, Saudi Arabia

ABSTRACT

The fact that almost every person owns a smartphone device that can be precisely located is both empowering and worrying. If methods for accurate tracking of devices (and their owners) via WiFi probing are developed in a responsible way, they could be applied in many different fields, from data security to urban planning. Numerous approaches to data collection and analysis have been covered, some of which use active sensing equipment, while others rely on passive probing, which takes advantage of nearly universal smartphone usage and WiFi network coverage. In this study, we introduce a system that uses WiFi probing technologies aimed at tracking user locations and understanding individual behavior. We built our own devices to passively capture WiFi request probe packets from smartphones, without the phones being connected to the network. The devices were tested at the headquarters of the research sector of the Elm Company. The results of the analyses carried out to estimate the crowd density in offices and the flows of the crowd from one place to another are promising and illustrate the importance of such solutions in indoor and closed spaces.

## 1 Introduction

In a world in which a large majority of people carry smart devices everywhere and smart buildings are becoming a new standard, there are far more opportunities to collect spatial data and use it to enhance environments and experiences Lee et al. [2021]. By accurately tracking a large number of devices in real time, researchers can gain valuable insights into behavioral patterns and identify hidden trends that would be difficult to recognize without this form of data. WiFi data represent a readily available source that can be easily tapped into, and therefore the development of advanced techniques that ensure efficient capturing with high precision is currently a priority. If successful, the current wave of studies targeting WiFi probing have the potential to radically change digital systems, as we know them today, and unlock the next level of artificial intelligence Sharma et al. [2021], Arasteh et al. [2016].

While studies on sensor-based location probing have only recently commenced, numerous studies from this field exist that deserve attention. Wang et al. Wang et al. [2013], Li et al. [2020a] suggested one of the first systems for the use of WiFi location probing in search and rescue operations, whereas the authors in Handte et al. [2014] experimented with estimating crowd density using similar concepts. Studies by Piccialli et al. [2019], Dogan et al. [2019] attempted to analyze human behavior based on Bluetooth signals from mobile phones with various degrees of success. Another study based on WiFi location probing was also proposed by Lesani and Miranda-Moreno [2018], wherein people were at large events, and Yin and Chi [2021] attempted to track people's movement in an urban environment. The authors in Kontokosta and Johnson [2017] also relied on WiFi data to analyze the behaviors of city dwellers, demonstrating how this technology can be used in a broader context. More specific applications of WiFi probing include studies by Petre et al. [2017], who analyzed human mobility and tracked crowd behavior during a festival. While all these studies demonstrate the viability of device tracking, there are many unresolved issues that need to be resolved through further studies.

In this study, an instrumental model that deals with the data analysis aspect of WiFi probing is proposed, which introduces methods to compare the trajectories of individual users, discovers trends from their interactions, and enables a more detailed analysis of group behavior. Understanding user journey behavior will help in several domains such as commercial, health, and crowd management. The proposed approach comprises two main components: hardware and software. Wi-Fi probe sensing hardware used in this model involved two custom-made devices, each containing a number of components such as WiFi module, LDO voltage regulator, nano Pi2 with Linux OS, and LED's. The main difference between them was the mode of network connection, with one device using LAN network and the other connecting to a wireless network. A back-end server supports the system, with MQTT protocol used for communication between the probing devices and the server. Software aspect of the system included two separate layers – locally embedded software within the device serving as the interface, and cloud-based applications that are hosted on the server. The embedded layer consists of a bash script that runs when the system is booted, and Python script that allows the user to control the device. Meanwhile, data collection service connects to the server and archives the incoming data, while density calculation service provides estimations of crowd density in real time using data from the server. Another application uses historical location data to determine people flow and map out user journeys. Finally, data analytics application allows for easy visualization of data and its re-organization through a user-friendly dashboard, and facilitates proactive setting of density limits for a given space.

The remainder of this paper is organized as follows. In Section 2, we provide a brief background of the most important concepts in this field. Section 3 describes the hardware design, Section 4 describes the experiments conducted to understand the behaviors of smart phones, Section 5 describes the embedded software, Section 6 discusses the cloud services, Section 7 describes our density estimation and journey, and Section 8 concludes the study.

## 2 Preliminaries

In this section, we provide general preliminaries of the different methodologies that can be used for people counting and user-movement tracking tasks.

### 2.1 People counting approaches

The need to manage indoor crowds is an increasingly important requirement in many fields. This domain is served by a wide range of technologies, including the latest smart tools. Several technologies can be used to track individuals and to count the total number of people in a given area. The most significant approaches include the computer-vision-based and WiFi sensing methods.

#### 2.1.1 Computer-vision-based methodology

Computer vision-based crowd counting has been growing in popularity in recent years, owing to its extraordinary accuracy. The main drawback of this system in practical deployment is its extremely high computing power requirement,
which makes it difficult to implement in large areas. An additional complication arises from the fact that face registration and recognition are the preconditions for tracking the flow of the crowd, raising all kinds of privacy-related, financial, and logistical issues.

#### 2.1.2 WiFi and probe packets

WiFi technology allows digital devices to send different types of frames to the WiFi packets. There three main categories are data, control, and management frames. Each frame can have different sub frames. Data frames are primarily used for data transmission, and control frames serve for channel acquisition, carrier sensing maintenance function, and acknowledgments, whereas management frames are used to join or leave a WiFi network and move association from one WiFi network to another WiFi network.

We performed a detailed study of WiFi probe packets classified as management frames. A mobile phone continuously broadcasts WiFi probe request packets to connect to a WiFi access point (AP). These probe packets are sent even if the mobile phone is already connected to a WiFi access point. This enables the device to remain connected to the WiFi network if there is a problem with the internet connectivity of that AP, or if the user moves away from the nearest AP. The WiFi probe packet mechanism allows the device to provide continuous Internet connectivity to the user.

Every mobile phone enabled with WiFi has a unique identifier called the MAC address. The MAC address is unique to each device and is created by the manufacturer to identify each product. Every WiFi probe packet contains the MAC address of the sending device. .

#### 2.1.3 Table abbreviations

In this section, we list some of the terms and abbreviations used in the study, as shown in Table 1.

Table 1: All Used Abbreviations

| Abbreviation | Stands for |
|---|---|
| LDO | Low-Dropout |
| POE | Power over ethernet |
| AP | Access Point |
| LAN | Local area network |
| LED | Light emitting diode |
| USB | universal serial bus |
| UART | universal asynchronous receiver-transmitter |
| IC | Integrated circuit |
| SSID | Service Set Identifier |
| MQTT | Message Queuing Telemetry Transport |
| MAC | Media Access Control |
| WiFi | Wireless Fidelity |
| OUI | Organizationally Unique Identifier |
| IP | Internet Protocol |
| RSSI | Received Signal Strength Indicator |
| UWB | ultra-wideband |
| RF | radio frequency |
| JSON | JavaScript Object Notation |

## 3 Literature review

Because WiFi probing is used in various contexts in the reviewed literature, many different methodological concepts have been introduced by researchers. The scope and environment of a project drive the selection of methods; for example, studies focused on indoor spaces logically have different requirements, such as large, open-space tracking projects that may cover areas as wide as Riyadh Cheng et al. [2021]. However, there is a sequence of operations that is common for most studies from this group, with several clearly defined stages. Data collection is the first task, and researchers choose an indicator such as a probe request to a mobile AP or MAC address randomization to identify the current location of each device or user. Passive and active scanning techniques can be deployed, which can either track requests directed to a specific SSID or capture all AP requests Zhou et al. [2020], Pu et al. [2021], Cheng et al. [2021].

The information collected during this stage can be divided into spatial, temporal, and semantic information, giving researchers many possibilities for interpretation and pattern detection. It is also possible to identify the manufacturer of a mobile device by analyzing its MAC address. After the data are collected, they are typically aggregated to create a dataset for further analysis. In studies focusing on occupancy estimation, aggregated numbers can be verified by manual observation Vattapparamban et al. [2016], Ciftler et al. [2017], Pereira et al. [2019], Azimi and O'Brien [2022]. Pre-processing operations also include cleanup of redundant requests and removal of devices that appear stationary, as well as interpolation for time periods. User identities may also be obscured by replacing the actual MAC addresses in the public dataset. The next step involves location detection for specific times, and this step is nearly universal, regardless of the purpose of the study. However, some studies are predominantly concerned with device counting and tracking the number of people in a certain zone or location Li et al. [2018], Singh et al. [2021], Alo et al. [2022].

However, studies aimed at analyzing movement track more complex spatio-temporal patterns indicative of a person's activity Traunmueller et al. [2018], Gu et al. [2021], Uras et al. [2020], Zhou et al. [2020]. Such studies typically create trajectories between WiFi APs, which may require the use of complex statistical operations, such as K-means clustering, density-based clustering, or hidden Markov models. In some cases, graphs can be constructed or trajectories can be superimposed over geographic maps or building floor plans Traunmueller et al. [2018]. Another key focus is describing interpersonal interactions, which can be understood through the construction of “social networks” comprising individuals who are regularly meeting in various locations and at various times and might be connected to others in a variable degree of separation Wang et al. [2017]. These connections can be deduced based on trajectory similarity and direct overlaps Wang et al. [2017], Li et al. [2021]. From this stage, it is possible to conduct data mining and search for intimate relationships between individuals, effectively turning simple location-based information into semantic links that can be used to classify users into different groups Zhou et al. [2020], Redondi and Cesana [2018]. Interpretation of WiFi data always requires a contextual understanding of the observed phenomenon; for example, individuals within an educational institution will interact with each other in different ways than people in a public setting, which must be considered when attempting to explain the findings Li et al. [2021]. Most methods for WiFi sniffing or probing were tested by their authors to a certain degree. The overall impression is that data collection and analysis techniques based on this concept can be very effective in the correct setting. Capturing the data was generally conducted with ease, and simple tasks such as occupancy detection or crowd estimation were completed very accurately, with over 90% of the devices correctly located within several meters. In such cases, it was often possible to compare findings with the ground truth based on direct observation to obtain accurate estimations. However, studies interested in large-scale group behavior often aimed to capture finer semantic relationships between subjects. Because many of the findings were qualitative, it was more difficult to verify how effectively these methods accomplish their objectives Uras et al. [2020], Li et al. [2020b], Retscher [2020].

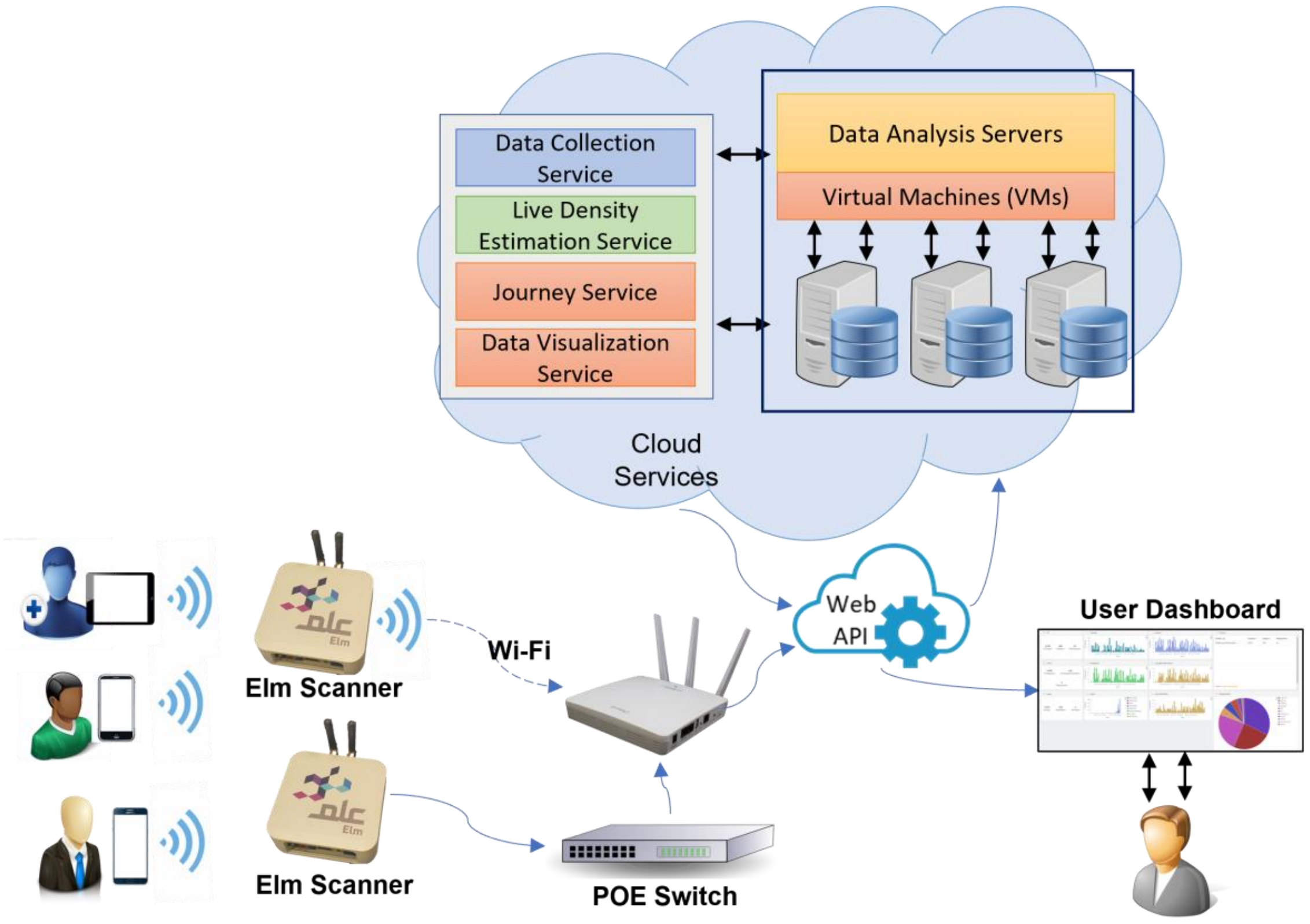

Figure 1: Proposed system architecture

More complex interactions are typically modeled on small samples of just a few hundred users, and should thus be seen more as demonstrations of the concept’s viability than actual experimental confirmations Li et al. [2021], Yang et al. [2021]. Some researchers have attempted to quantify the key elements of their studies, including trajectory inference, which is a crucial step for tracking movement and understanding patterns of social activity Yang et al. [2021]. Although the tested methods were shown to be operable and reasonably successful under the test scenario, it is difficult to see any of them as complete and ready for commercial exploitation. Hence, the overall achievements of these studies can be described as intriguing and inspiring, but hardly definitive in any sense.

# 4 System architecture

The proposed system architecture is illustrated in Figure 1. The system comprises three main components. The first component is the hardware component, which contains all the devices developed and used in this study. Another key element is the cloud service component, which includes online services developed for data collection, analysis, and presentation to the user. Finally, there is an embedded software component, which is concerned with the software that is built into the device and serves to connect the hardware, software, and end user. All three components are described in the following subsections.

## 4.1 Hardware component

Owing to the lack of WiFi sensing hardware off the shelf, we decided to develop our own device that would be tailored to our custom needs. Considering the requirements of this device and the expected deployment scenarios, two types of sensing devices were built, as shown in Figure 2. Each device had a dedicated WiFi module, which was used for scanning only. There were two options for Internet connectivity: with one device using a WiFi module and connecting to an AP Figure 2. This device obtained input power from a 5 V wall adapter via a USB cable. The module contained an LDO for the voltage regulator, two WiFi modules, nano Pi2, LEDs, and a USB to UART IC.

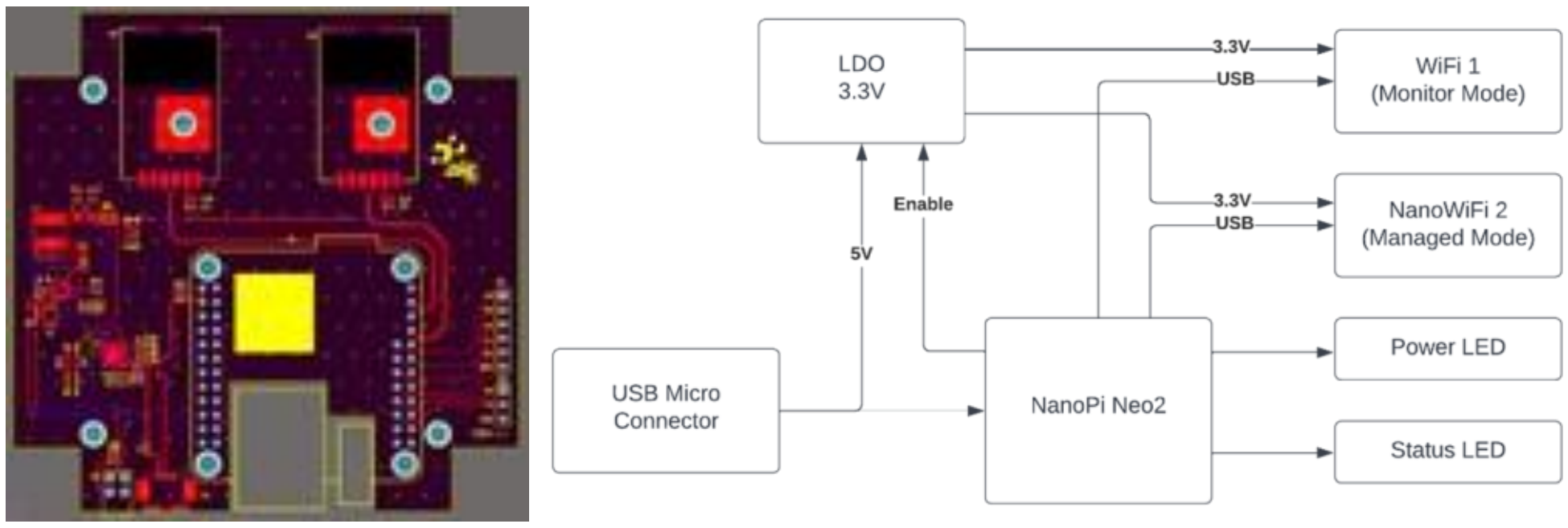

Figure 2: Developed hardware device powered by 5 V USB adapter.

The other device used power over ethernet (POE) to connect LAN networks, as shown in Figure 3. This version contained a POE splitter, LDO voltage regulator, WiFi module, nano Pi2, LEDs, and USB-to-UART IC. Nano Pi2 technology supported the Linux operation system at the core of both devices. It had an Allwinnder H5 Quad-core64bit cortex A53 CPU with 512 MB DDR3 RAM, along with other interfaces such as GPIOs and USB. This provided more processing power and the flexibility necessary to perform certain tasks on the edge. Figure 4 shows the assembled WiFi sensing device that contains two WiFi modules.

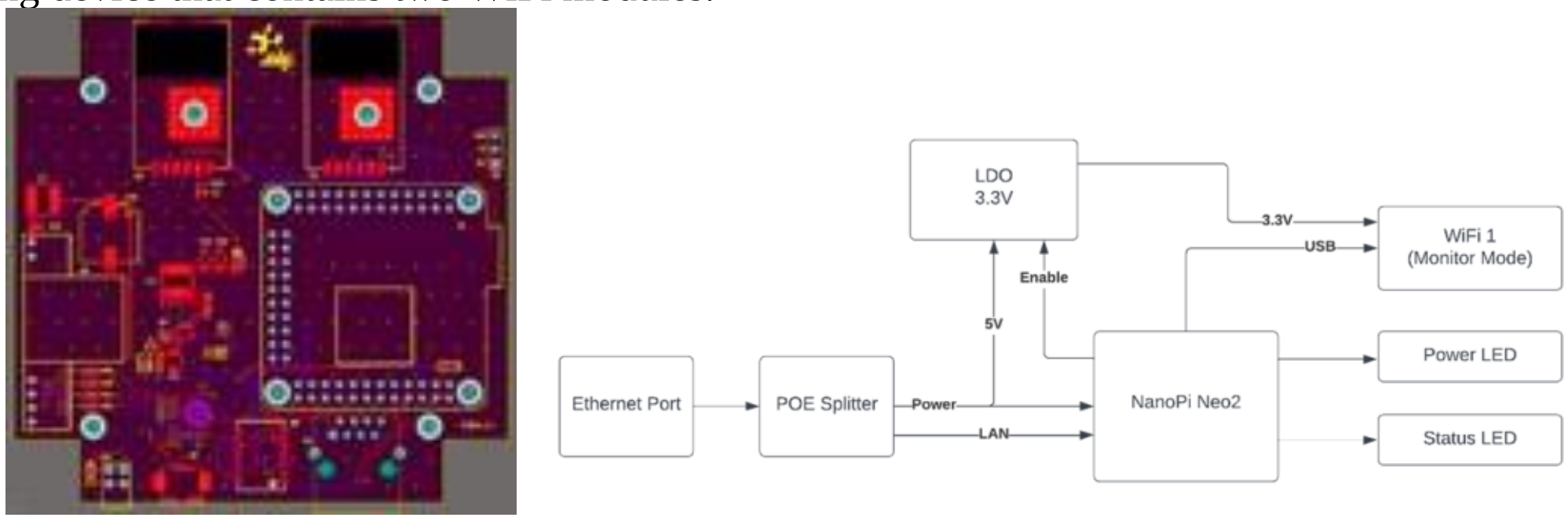

Figure 3: Developed hardware device powered by POE

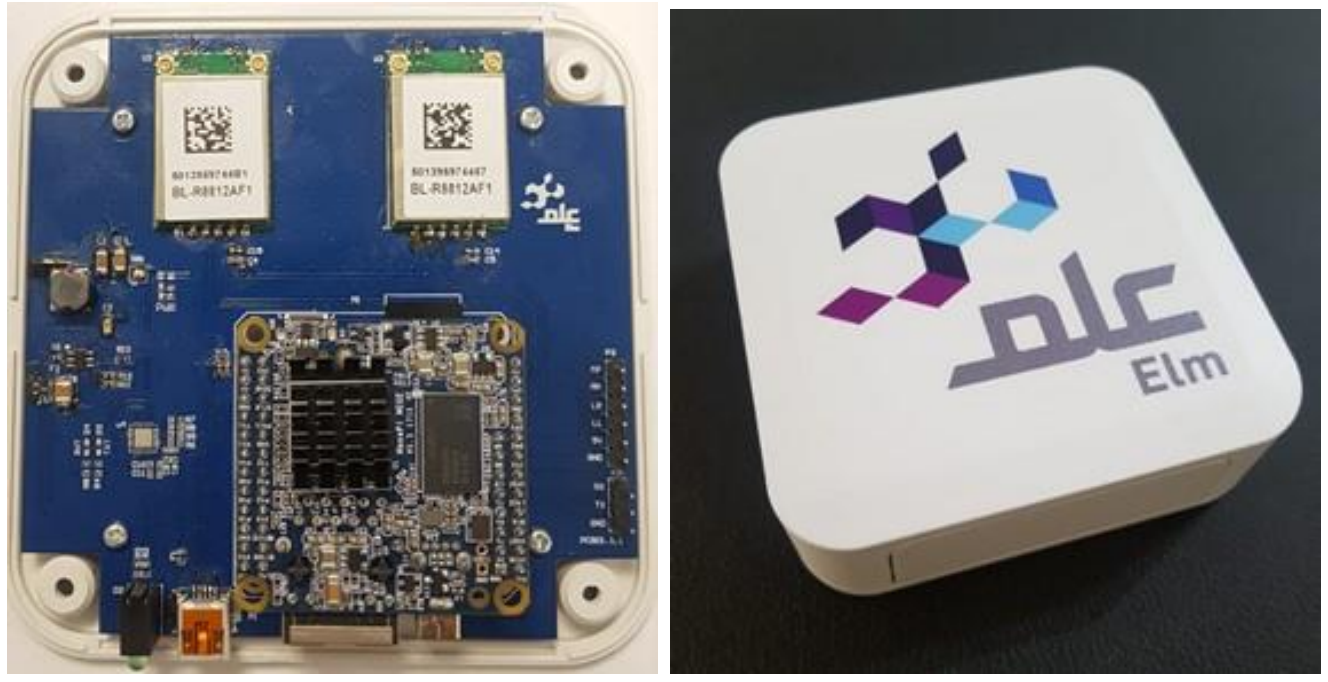

Figure 4: Developed and assembled device

#### 4.1.1 Embedded software

To scan for WiFi probe packets, the WiFi interface needed to be set to the monitor mode. This was accomplished by using a bash script that converted the WiFi mode from managed to monitored. As the mode of the interface was not permanently saved, this script ran every time the system booted. After this script successfully changed the mode, the main Python script that performed the scanning was launched. Immediately after launching, it established a connection to the server through MQTT. Every time it connected with MQTT, it sent a birth packet on a log topic and set the last will packet. This packet was sent to the server, and if the connection with the device was lost, this message was published, indicating that the device had stopped working. All packets contained the MAC address as a unique identifier. The birth packet also contained the version of the software the device was running and its local IP address that was used for maintenance, if required. The initialization process is illustrated in Figure 5.

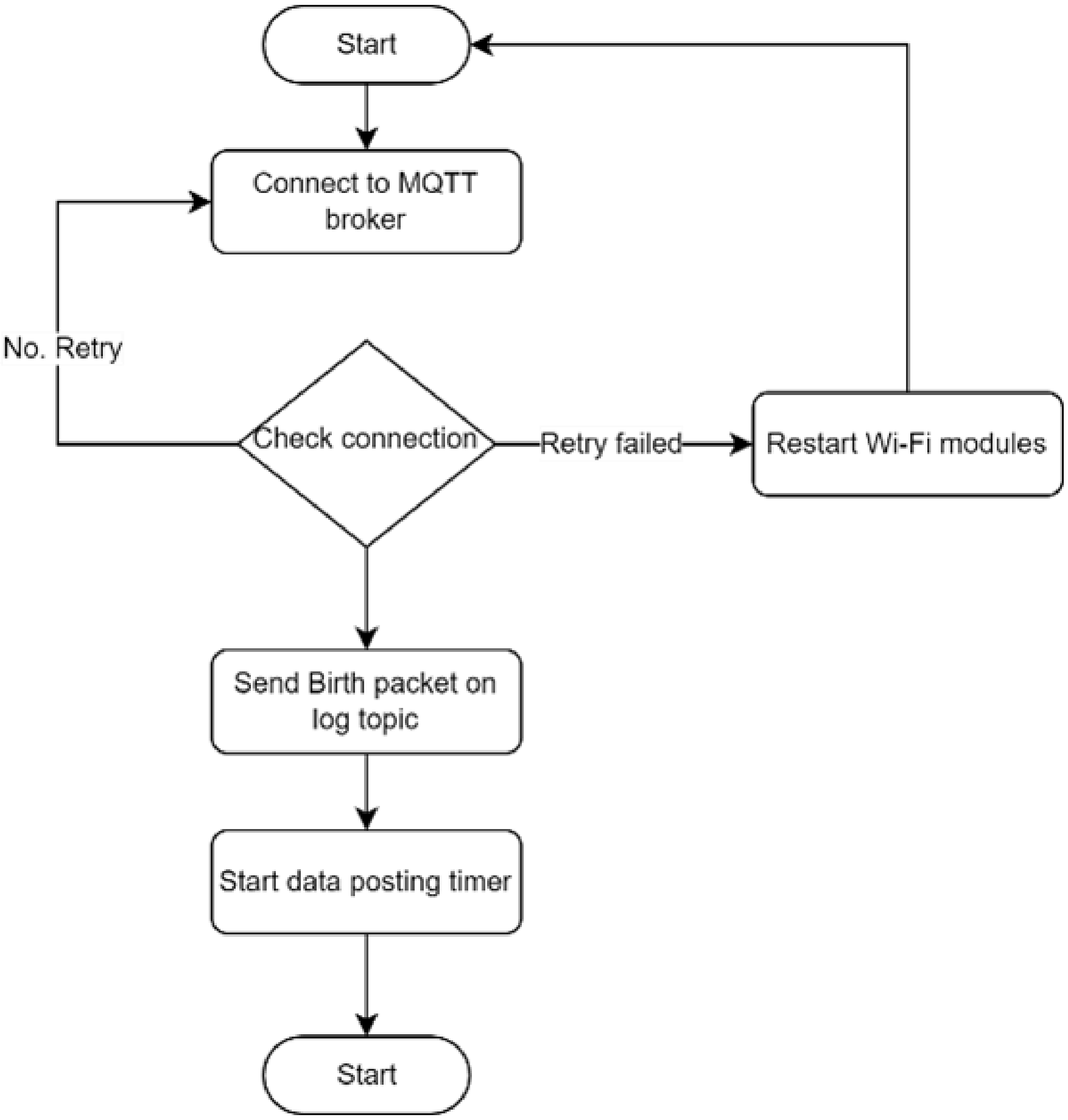

Figure 5: Initialization flow chart for the main script

Once the MQTT connection was successfully established, the device started scanning for the packets. The device scanned only WiFi probe packets and ignored all other packets. A WiFi package contained the MAC address of the device, the received signal strength, SSIDs of previously connected networks, information elements, and vendor-specific details. With the help of updated organizationally unique identifier (OUI) we learned the manufacturer of the device behind each MAC address. Based on the OUI information, vendors that were not known for manufacturing mobile phones were ignored. This information, along with the time when the packets were received, was temporarily stored inside the device until the packet transmission interval was reached.

Once the data were received, they were preprocessed to reduce the network overhead and optimize the packets, as depicted in Figure 6. Based on the conducted experiments, multiple probe request packets were received for each probe request interval, and most of the time, these packets had very slight changes in content. Small changes were routinely ignored; however, all the major changes were notified. To reduce the packet size, md5 HASH of the information element data and vendor-specific data were generated and stored in the device. The device maintained the stored data until the posting interval. Once the posting interval was reached, the device sent compiled data in JSON format and published it on the MQTT data topic.

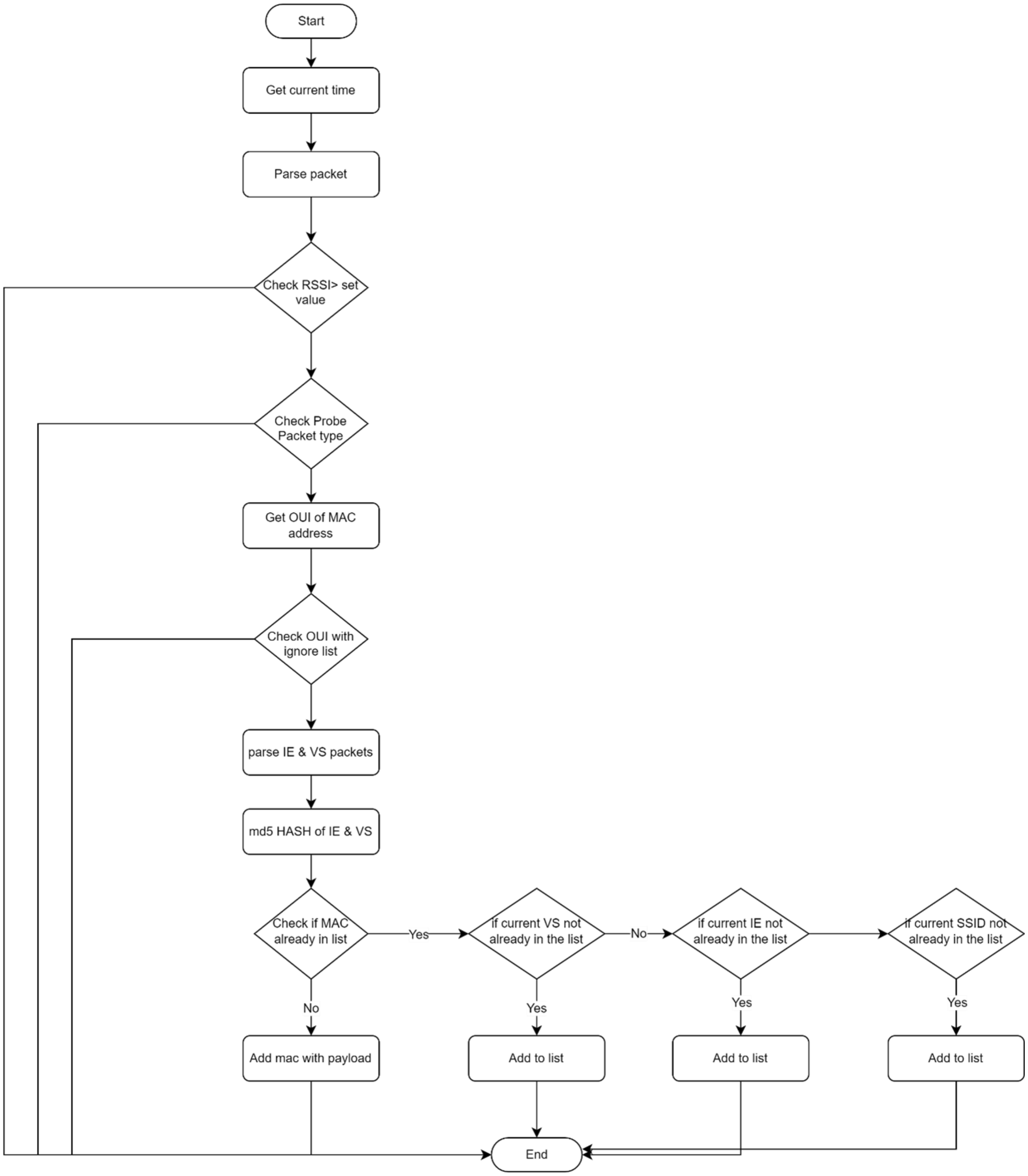

Figure 6: Preprocessing the data after receiving the WiFi probe request packet

### 4.2 Cloud services component

Several cloud-based services were developed that are useful for tasks such as collecting data from devices, calculating the density in an area at a particular time, monitoring the user's journey, and performing advanced data analysis and visualization. In addition, the mosquito MQTT server was deployed to communicate with the devices. Details of these services are explained in the following sections.

#### 4.2.1 MQTT Server

For the communication of the devices with the backend server, MQTT standard for IoT messaging was used. It is a lightweight messaging protocol. Each client in MQTT connected to the MQTT broker. Clients published messages or data using a topic. Services that needed to use the data subscribed to that topic and received the data in real-time once it
was published. This rendered the system scalable and reliable.

#### 4.2.2 Data collection service

A data collection service was used to archive the raw data arriving from the devices. This service connected to the MQTT server, and whenever there were data from the available devices, it was parsed and stored in the database. The stored data contained the date, time, and MAC address or ID of each scanned device. The stored data were used to perform analytic tasks and recalculate the density if there were any updates in the density calculation algorithm. This process is illustrated in Figure 7.

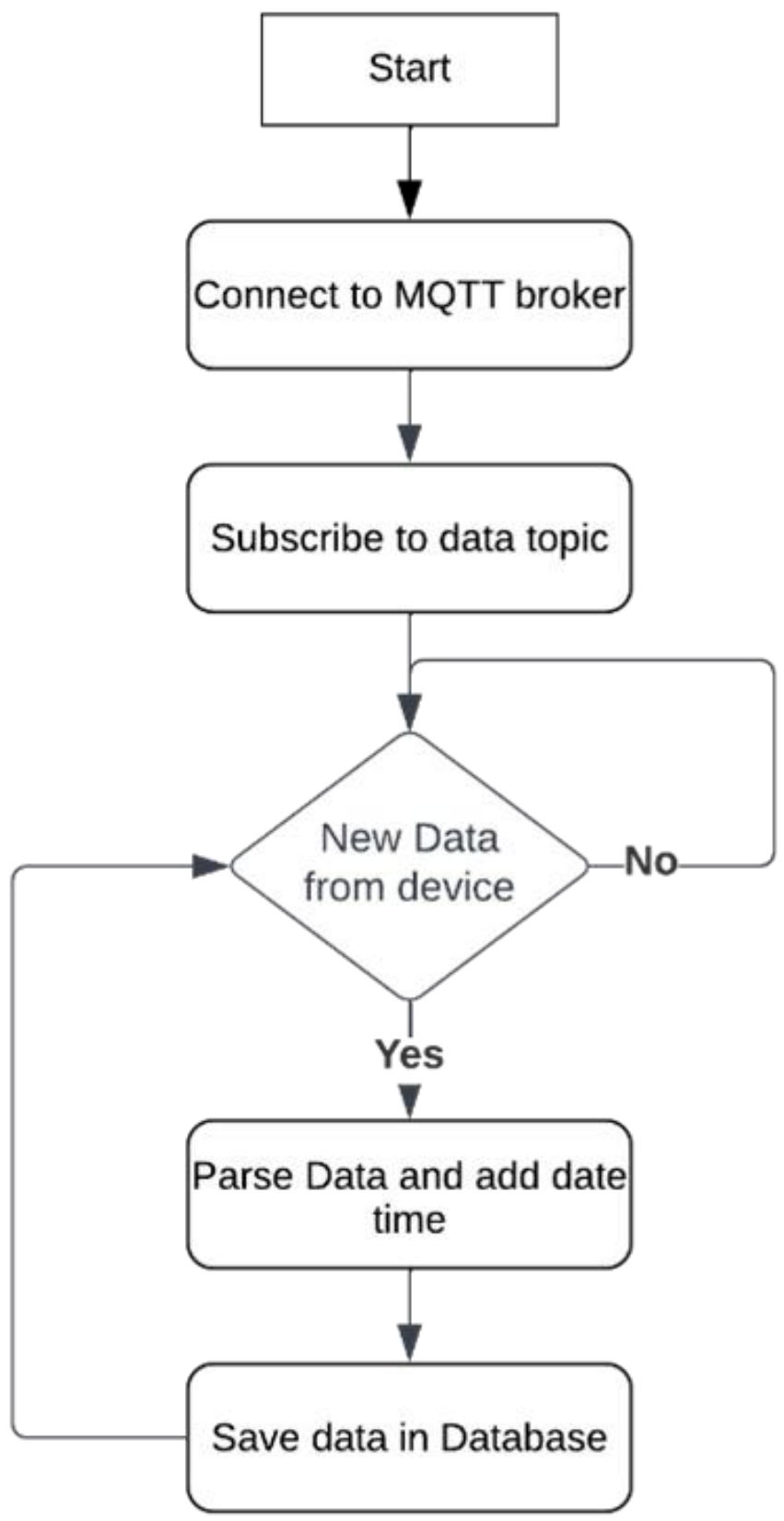

Figure 7: Process of receiving data on cloud server and storing it in database

#### 4.2.3 Density calculation service

The density calculation service worked with sensor data in real-time. The service first connected to a database and collected user information for a specific company. It then recognized all the buildings associated with a company and all the unique ID scanners in that building. To receive real-time data, it connected to the MQTT broker and subscribed to relevant topics. Each device consolidated the data (as mentioned above) and published them on MQTT within its posting interval.

Upon receiving the data from the device, a dictionary was maintained for all probe packets with the time when they arrived. The MAC address of the probe request packet was the unique key. Every time a set of probe packets was received from the scanning device, the time the last packet was received was updated. After a configurable interval of 60 s, the time at which the last packets from the MAC were received was compared with the current time. If the time difference was greater than 240 s or 4 min, the packet was removed from the list. During the experiments, it was observed that, even if the MAC address was randomized, it still sent the same randomized MAC address of that phone in most cases.

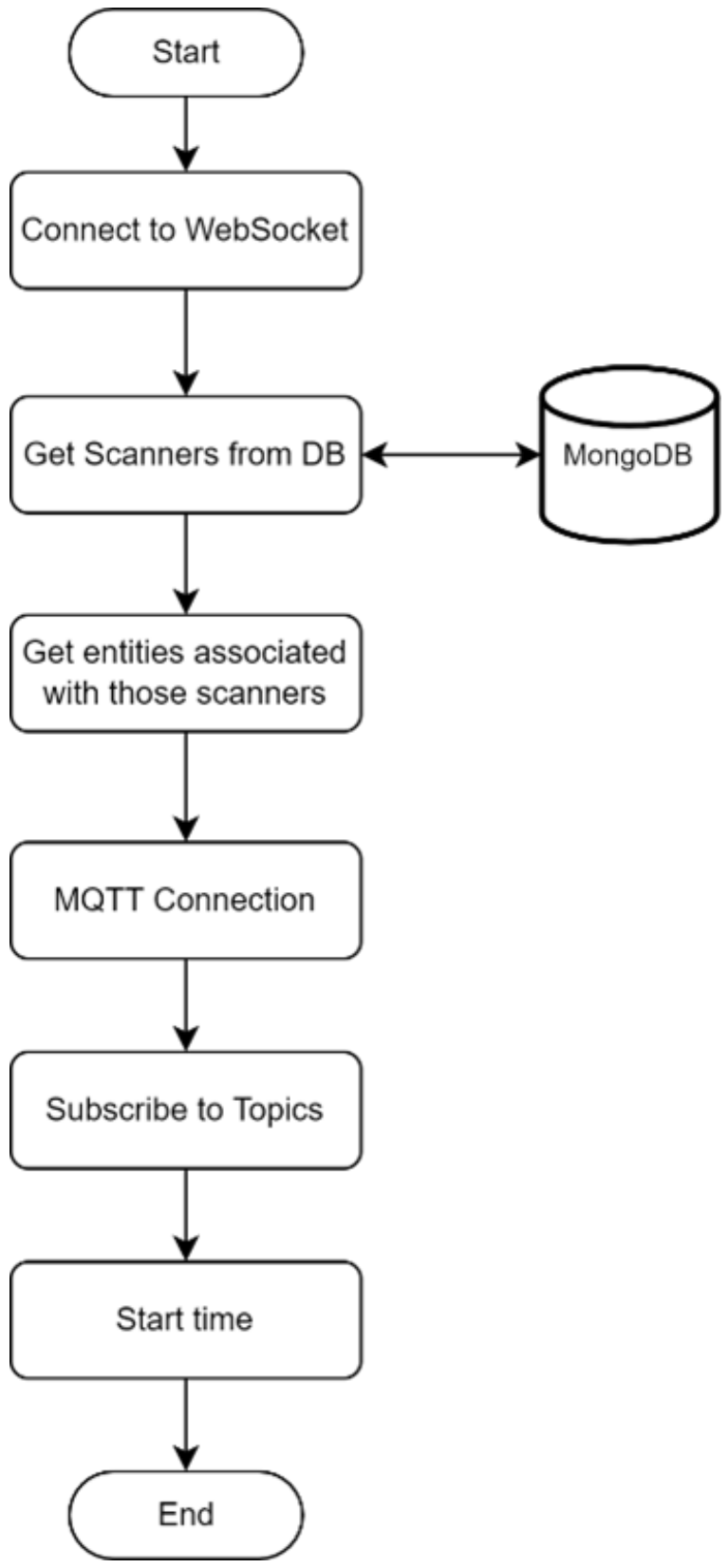

Figure 8: Initialization process of the service

The total number of leftover MAC addresses was counted for each scanner and sent over WebSocket for real-time updates on the dashboard. The count was also saved in the database for historical records of the number of people in that area. This combination of processing data at the edge and processing the remaining data in the cloud for density calculation provided an approximate count of people in the area near the scanner. Depending on practical concerns, further improvements can be made to the scanning devices to improve their accuracy by adding a directional antenna and setting up the latest RSSI filter.

#### 4.2.4 User journey

To obtain user journey estimations or flow calculations, all non-randomized MAC addresses were analyzed. The system checked all scanners in which the MACs were detected. Time-based detection was then performed to determine the temporal distribution of the input. In this step, the location where a particular MAC was first detected was noted and matched with the detection of that MAC by another scanner, consequently increasing the number of people in that direction. Another technique partially ignored the MAC address and focused on the information elements contained in the probe packets. Even when a mobile phone randomized the MAC address, it sent the same information element, which made it a reliable differentiator. However, there were some instances where smartphone models sent identical information elements, which made it difficult to track the flow/journey of such devices.

#### 4.2.5 Data analysis service

A dashboard was built to visualize and present the data. This dashboard had several components and was designed to consider different users and entities. Only a super admin could create an entity, but anyone with admin rights or even a normal user could add users to that entity. The admin could log in and create buildings, upload the floor map in any image format, place scanners on the map, and define the maximum density for that floor.

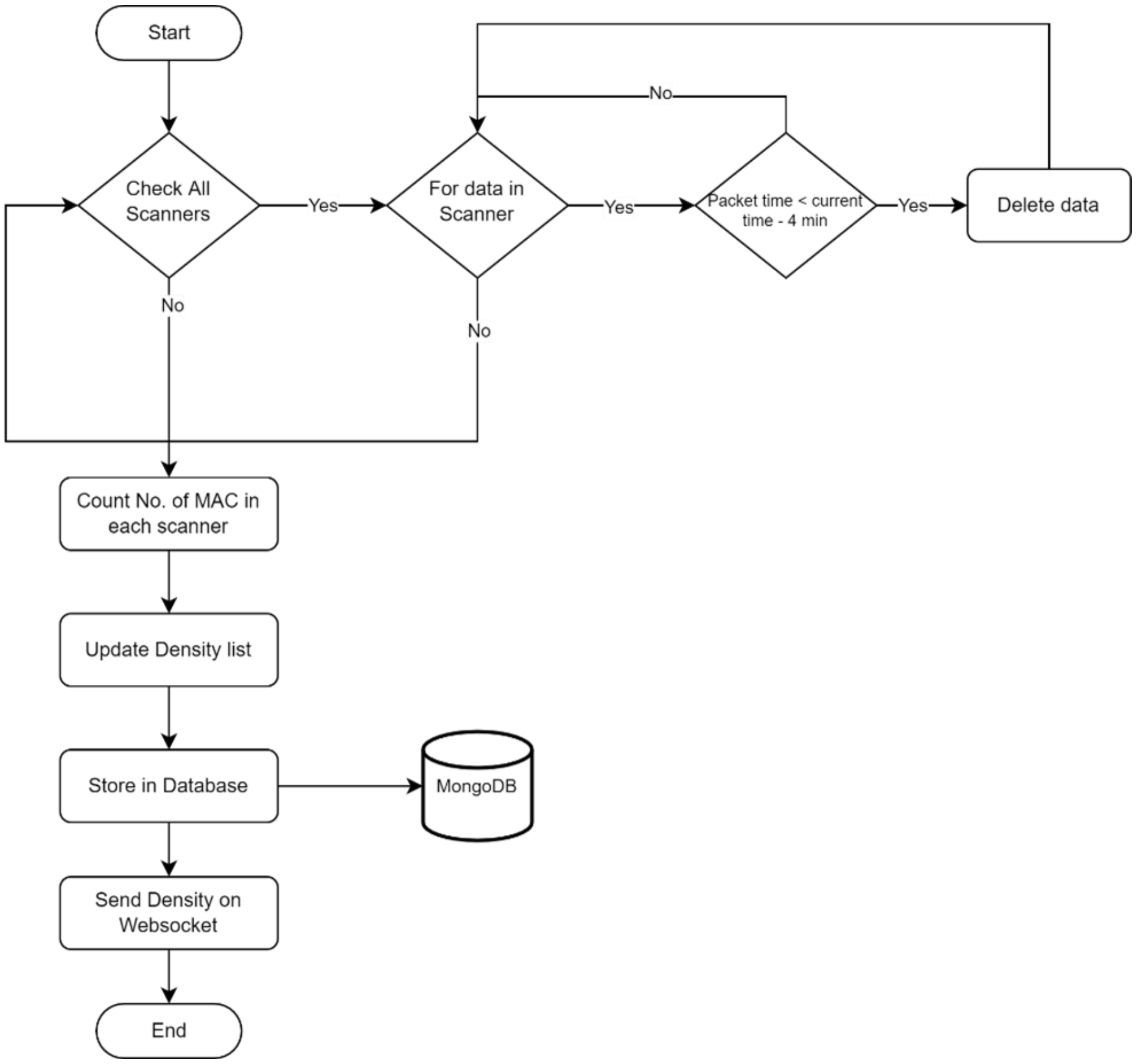

Figure 9: Density estimation process

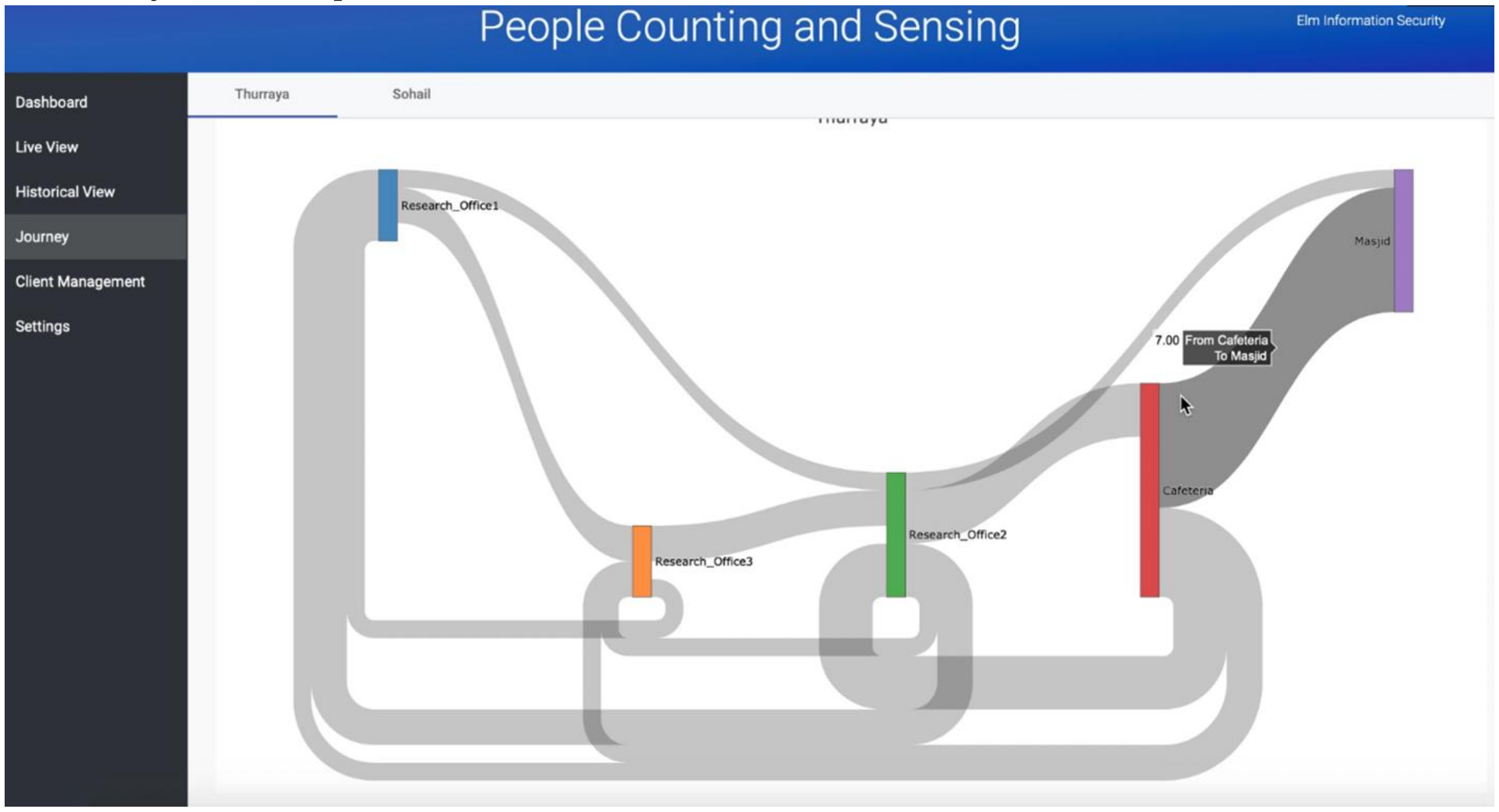

Figure 10: Visualizations of user journey

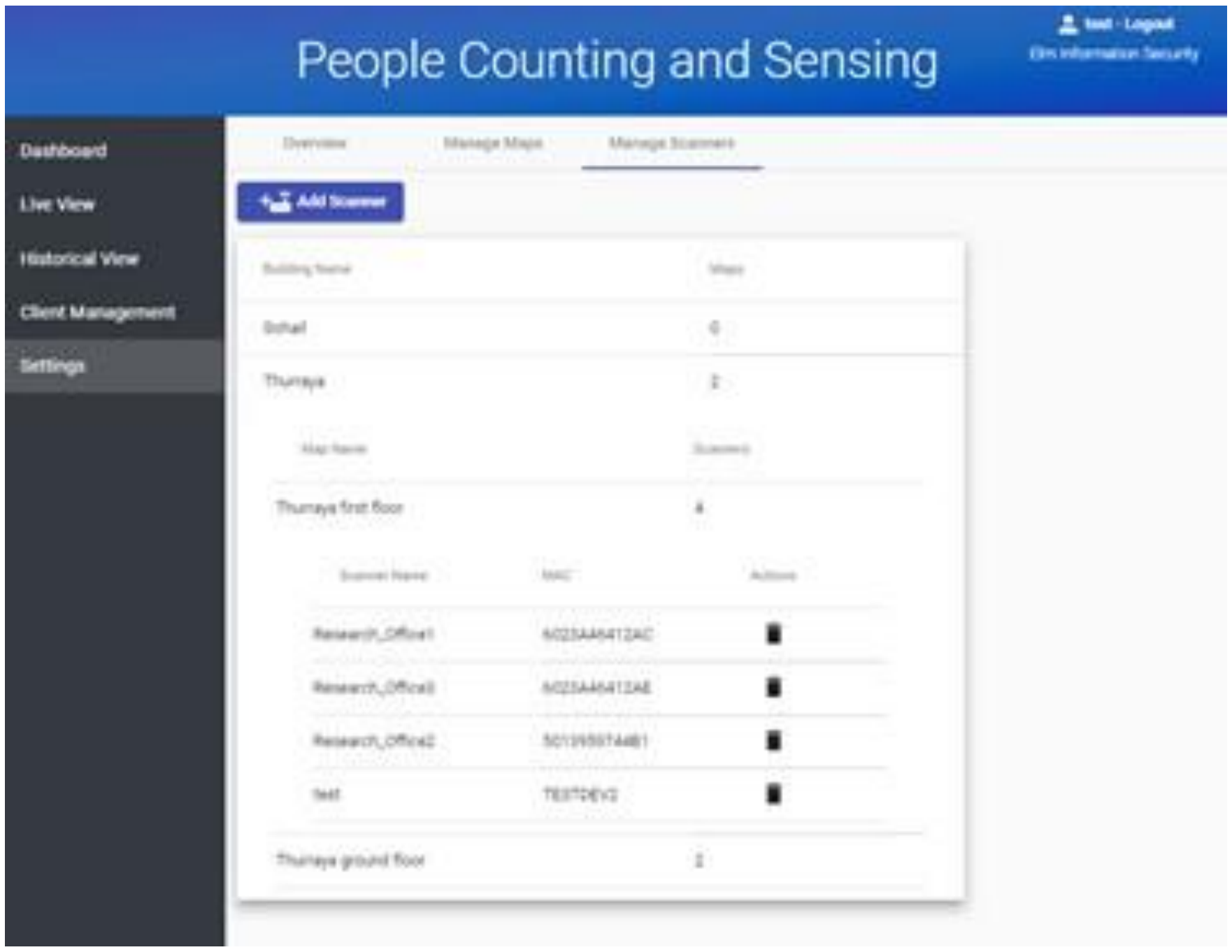

Figure 11: Admin portal to add floor map and devices

On the dashboard, users could view the density of people in the room with scanners in the form of line charts, as shown in Figure 12, or heat-maps, as shown in Figure 12. The user could select a date and view historical data in any of these two formats. Another feature of the dashboard is the journey of the people in the Sankey diagram, as shown in Figure 10.

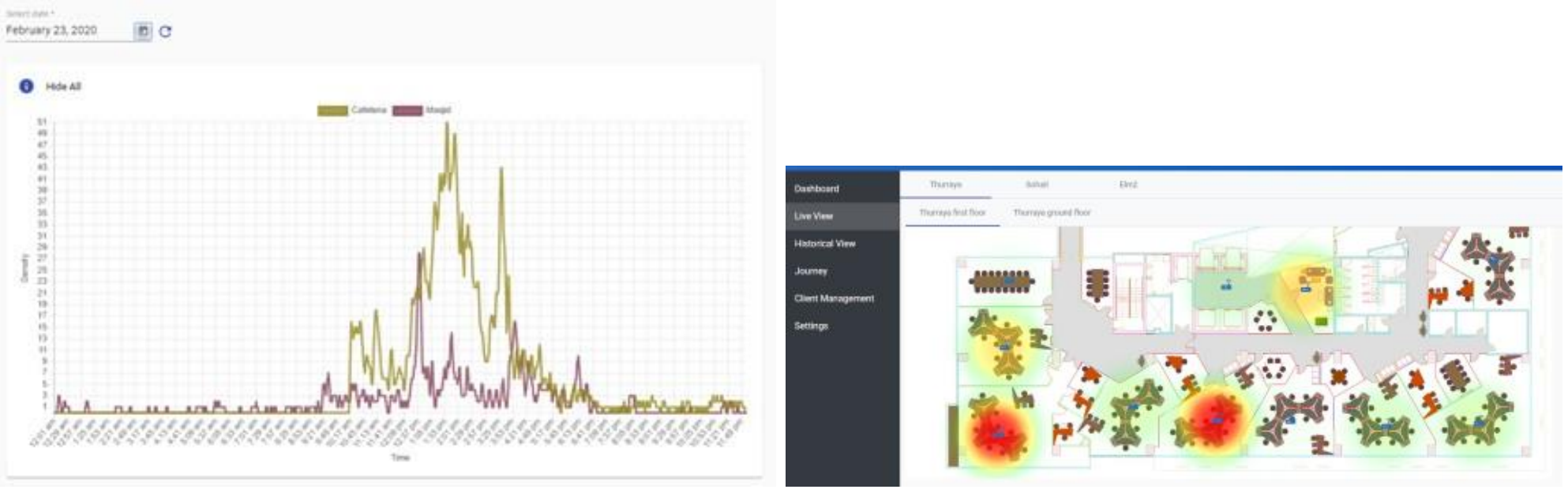

Figure 12: User dashboard

## 5 Experiment

To understand the characteristics of probe request packets from various mobile phones, experiments were conducted with different types of devices under isolated conditions. In these experiments, the WiFi USB dongle was connected to the Raspberry Pi, and the WiFi interface operating mode was changed to monitor mode in order to scan the packets. A Python script was created to interface with the WiFi dongle, scanned for all the packets, and then filtered out only WiFi probe request packets.

### 5.1 Test scenarios

In order to test the system in a controlled environment where there were no other probe packets, the Raspberry Pi was powered by a battery bank, and together with a mobile phone, scanning hardware was placed in a closed Faraday bag for an hour, as shown in Figure 13. The Faraday bag blocked all external RF communication and allowed the scanning hardware to scan only signals originating from the mobile phone kept inside it. To understand the behavior of WiFi on a mobile phone, several scenarios, as listed in Table 2, were tested.

Table 2: All Used Mobile Phones

| Mobiles | WiFi Status | Screen | Test Duration |
|---|---|---|---|
| Samsung S7 | Off | Off | 60 minutes |
| Samsung S7 | On | Off | 60 minutes |
| Samsung S7 | On | On | 60 minutes |
| iPhone 6 | Off | Off | 60 minutes |
| iPhone 6 | On | Off | 60 minutes |
| iPhone 6 | On | On | 60 minutes |
| Samsung J5 | Off | Off | 60 minutes |
| Samsung J5 | On | Off | 60 minutes |
| Samsung J5 | On | On | 60 minutes |
| Xiaomi Mi Note 3 | Off | Off | 60 minutes |
| Xiaomi Mi Note 3 | On | Off | 60 minutes |
| Xiaomi Mi Note 3 | On | On | 60 minutes |

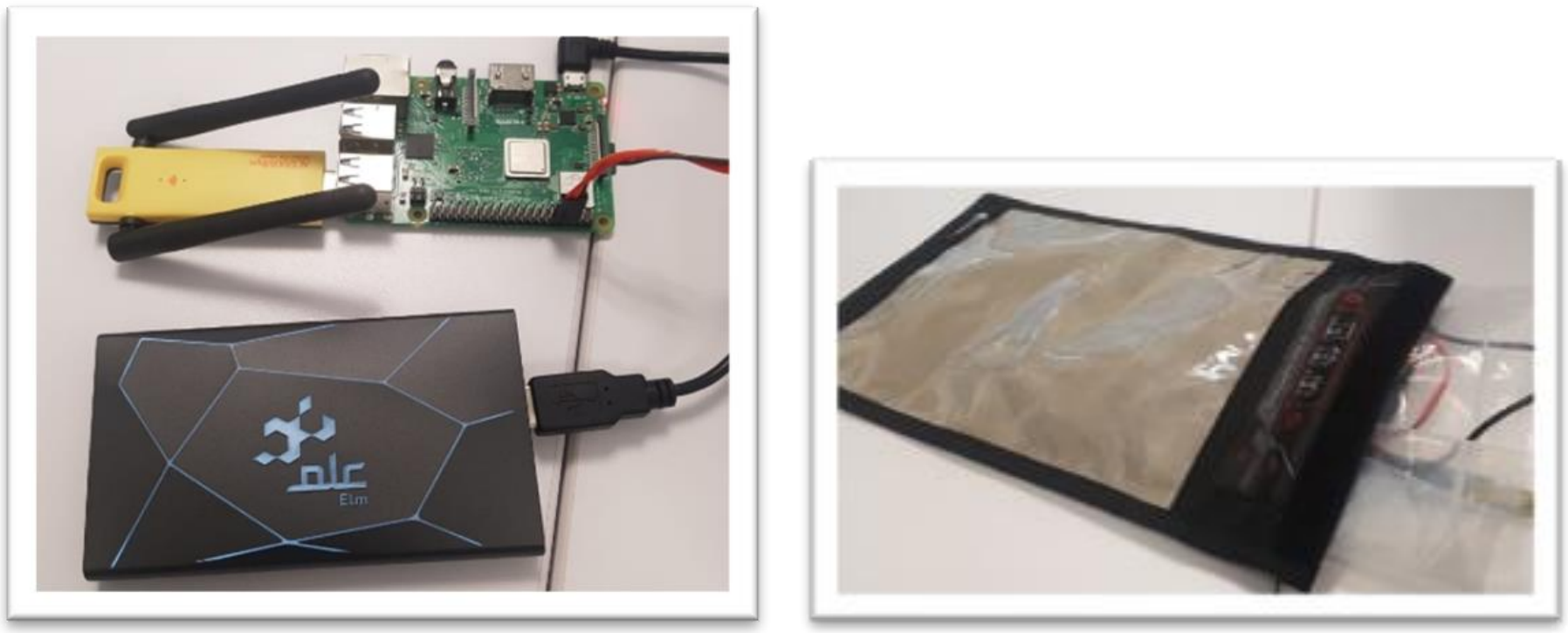

Figure 13: Scanning hardware was kept in a Faraday bag and closed for an hour.

As shown in Table 2, the mobile phones tested were Samsung S7, iPhone 6, Samsung J5, and Xiaomi Mi Note 3. To understand how each mobile phone acted with and without WiFi access and screen activation, the following scenarios were tested: 1) Mobile phone had WiFi access disabled 2) Mobile phone had WiFi access enabled and its display turned off 3) Mobile phone had WiFi access enabled and its display turned on

The captured data contained the exact times at which it was captured, MAC address, RSSI value, SSIDs in the information element, information element, and vendor-specific data. All these details were stored in a csv file inside the Raspberry Pi device, which was retrieved and analyzed at the end of the experiment.

### 5.2 Results

In all the figures below, the x-axis represents the time at which the experiment was conducted. The y-axis on the left shows the number of probe request packets recorded during each probing event. The y-axis on the right tracks the interval between the two probing events. Figure 14 provides a summary of the first scenario for iPhone 6S when the mobile screen was turned off. It was observed that the MAC address of the mobile phone was randomized, and that a total of 10 probing events were received in an hour. Figure 14 illustrates the second scenario for the same mobile phone, but with the screen active. With these parameters there was a significant increase in the frequency of probing events. Almost 54 probe events were received in an hour in total.

Figure 15 depicts the first scenario for Samsung S7 edge when the mobile screen was off. In this case, the findings included that the MAC address of the mobile phone was randomized, a total of 13 probing events were received in an hour. Figure 15 is for the same mobile phone, but its display was turned on, again yielding an increased number of 54 probing events in an hour. Here it was also observed that the mobile phone was sending probe packages with an almost fixed interval, however MAC randomization was still present. Another key observation was that throughout the duration of the test the information element remained constant even if the phone's MAC was randomized.

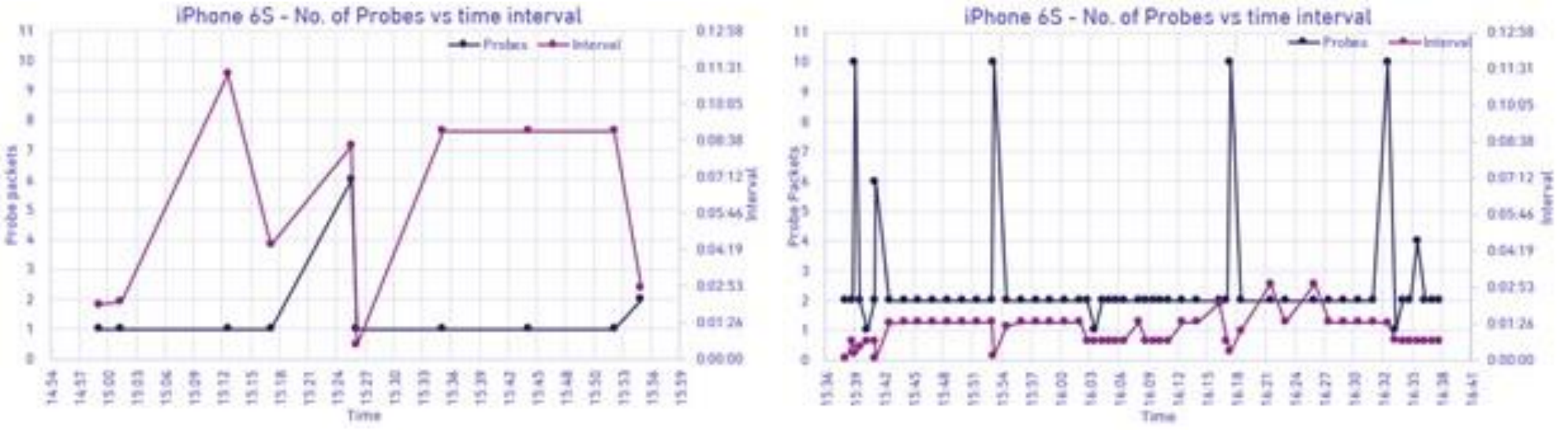

Figure 14: Output of the experiment conducted on iPhone 6S

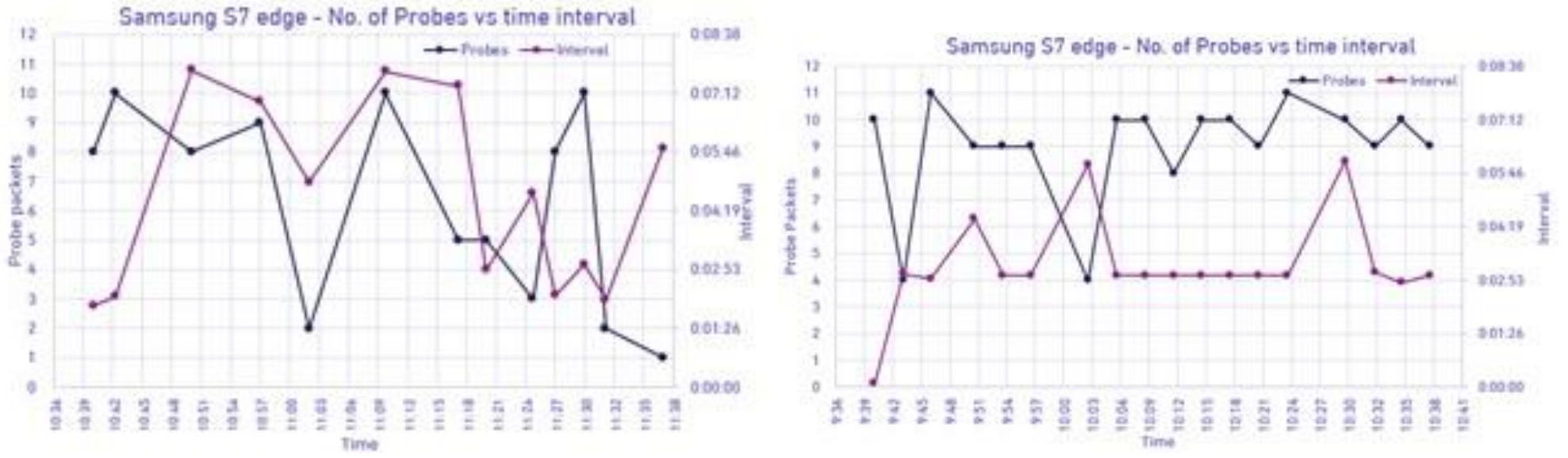

Figure 15: Output of the experiment conducted on Samsung S7 edge

Figure 16 presents data for the scenario #1 for Samsung J5, when the mobile screen was switched off. With this phone, the MAC address was not randomized, and only four probing events occurred within an hour. Figure 16 is dedicated to the second scenario for the same mobile phone with the difference that the screen was active. It was once again found that the number of the probing events spiked dramatically with change in the screen status. Nineteen probe events were received in an hour, a nearly 500% increase, while the interval remained similarly stable as for scenario #1.

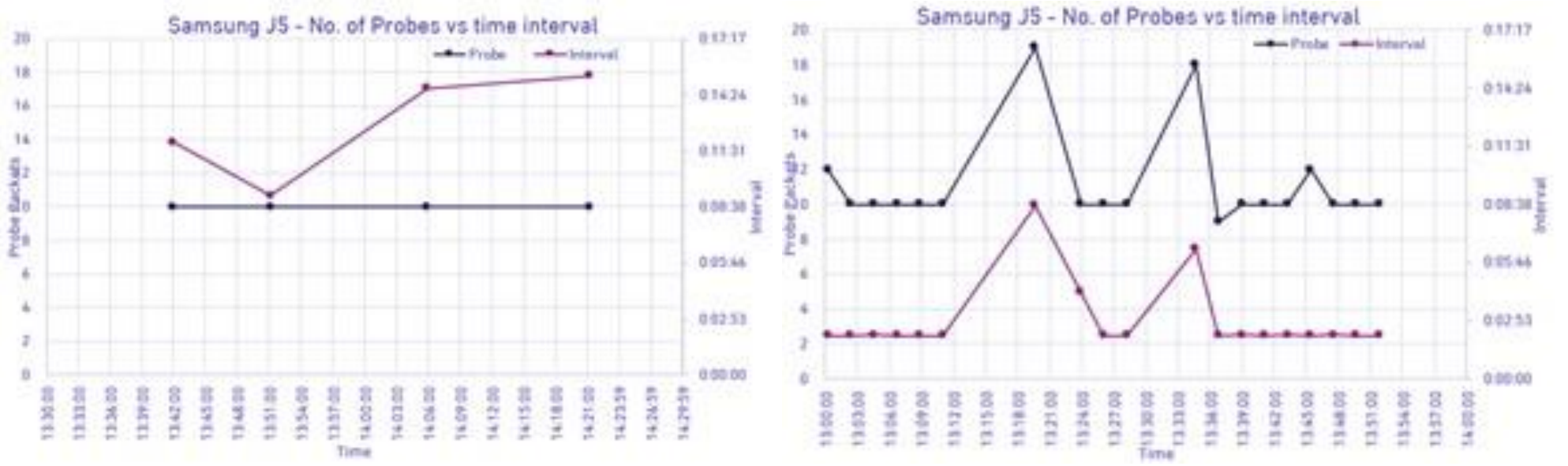

Figure 16: Output of our experiment conducted on Samsung J5

In Figure 17, the results for the first scenario using Xiaomi Mi Note 3 with a turned off display can be observed. This device did not randomize the MAC address and sent a total of 89 probe requests over the course of an hour, which was significantly higher than that for the other mobile phones tested. The sending interval was very regular, with nearly identical times elapsed between the two probes. Figure 17 presents the outcomes of the second scenario for the same mobile phone when the screen was active. Contrary to this trend, a large drop in the number of probing events was observed; however, a solid number of events were tracked compared with other mobile phones, at 24 per hour.

### 5.3 Discussion

The experiment concluded that no WiFi probe packets were broadcasted when WiFi was disabled and location services were off. Mobile phones broadcasted very few packets when WiFi access was not available but location services were enabled. Some phones have adopted MAC randomization for user privacy purposes; however, the majority of phones have not yet implemented this procedure. The interval between WiFi probe request events is not constant because they are essentially random. When a mobile phone was unlocked and its screen was turned on, it tended to send more probe request packets over the same time period. The transmission power of WiFi probe packets occasionally decreased when the screen was off, although this does not happen with all tested models and depended on the mechanisms implemented by the manufacturer. Probe packets contained SSIDs of previously connected networks, particularly when the SSID of an access point was hidden and had to be added to the phone manually. This happened because an AP with a hidden SSID did not advertise its presence, and therefore the WiFi-enabled devices sent probe request packets and waited for a response to make the connection. A mobile phone generally sends more packets when the screen is turned on. Data inside the information element do not randomize with each probe; however, it does change after a power cycle of enabling or disabling the WiFi interface, as well as a consequence of timing out after a couple of hours. Table 3 provides a summary of the results for all tested mobile devices.

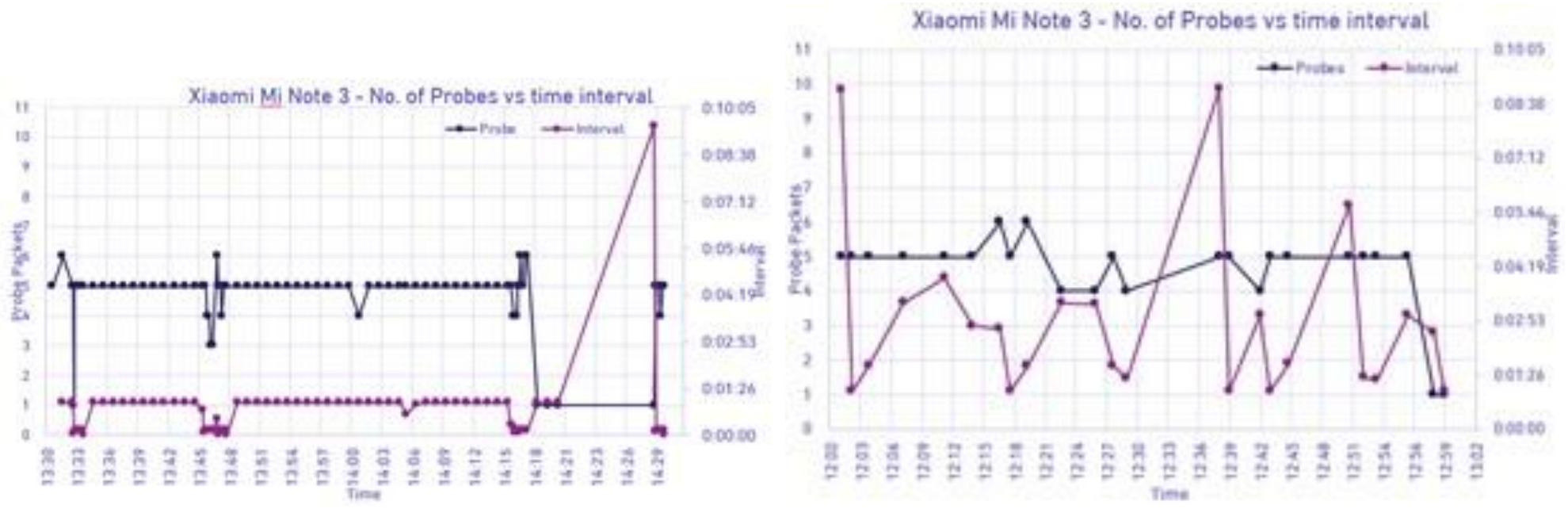

Figure 17: Output of our experiment conducted on Xiaomi Mi Note 3

Table 3: Summary of the results for all tested mobile devices

| Mobile Phone | MAC | Status | No of probes events/hr | Avg. No of probe packets | Probe packet interval |
|---|---|---|---|---|---|
| iPhone 6S | Randomized on every probe event | Display Off | 10 | 1 | Min = 33 s and Max = 11 m 15 s and Mode = 9 m |
| | | Display On | 54 | 2 | Min = 3 s<br>Max = 3 m<br>Mode = 45 s |
| Samsung S7 Android 8.0.0 | Randomized on every probe event | Display Off | 13 | 6 | Min = 2 m 9 s<br>Max = 7 m 46 s<br>Mode = 2 m 10 s |
| | | Display On | 18 | 9 | Min = 6 s<br>Max = 6 m 5 s<br>Mode = 3 m |
| Samsung J5 Android 6.0.1 | No Random | Display Off | 4 | 10 | Min = 9 m 12 s<br>Max = 15 m 22 s |
| | | Display On | 19 | 10 | Min = 2 m 8 s<br>Max = 8 m 36 s<br>Mode = 2 m 8 s |
| Xiaomi Mi Note 3 Android 7.1.1 | No Random | Display Off | 89 | 5 | Min = 1 s<br>Max = 9 m 29 s<br>Mode = 1 m |
| | | Display On | 24 | 5 | Min = 1 m<br>Max = 9 m 2 s<br>Mode = 1 m |

# 6 Conclusion

Ubiquitous presence of mobile phones opens some new possibilities for developing practical applications that wouldn't be possible just a decade or two ago. This paper describes an innovative WiFi sensing system based on a custom-built hardware configuration that is capable of capturing the WiFi probes containing MAC addresses emitted by smartphones in regular intervals, and sending them to the cloud for processing. The system was tested only under laboratory condition to determine some fundamental trends with probe signal capturing for devices from different manufacturers. In conclusion, we can say that, with the architecture that includes data preprocessing on the edge and the use of a cloud server for the remaining operations reduces the load on the network, as well as the typical packet size. With this complete end-to-end solution, we were able to achieve high accuracy for the task of estimating the number of people carrying mobile phones, with the requirement that the devices do not randomize the MAC address and stay near the point of interest for at least 2 to 3 min. With devices that randomize MAC addresses, we were able to achieve good accuracy for the same task. However, these results could change over time owing to changes in the design and features of the new smartphone models. Smartphone manufacturers are continuously trying to hide the identity of users by tweaking different WiFi probing mechanics. To upgrade the system, we recommend conducting similar experiments and upgrading the algorithms accordingly.

## Acknowledgments

This is a short text to acknowledge the contributions of specific colleagues, institutions, or agencies that aided the efforts of the authors.